\documentclass[twocolumn,,amsmath,amssymb]{revtex4}

\usepackage{graphicx}
\usepackage{dcolumn}
\usepackage{bm}
\newcommand{\etal}{\emph{et al.}}
\newcommand{\be}{\begin{equation}}
\newcommand{\ee}{\end{equation}}
\newcommand{\bfig}{\begin{figure}}
\newcommand{\efig}{\end{figure}}

\begin{document}      
\title{Thermal Hall conductivity in the frustrated pyrochlore magnet Tb$_2$Ti$_2$O$_7$
} 

\author{Max Hirschberger$^{1}$}
\author{Jason W. Krizan$^{2}$}
\author{R. J. Cava$^2$}
\author{N. P. Ong$^{1,*}$}
\affiliation{
Departments of Physics$^1$ and Chemistry$^2$, Princeton University, Princeton, NJ 08544
}

\date{\today}      
\begin{abstract}
In a ferromagnet, the spin excitations are the well-studied magnons. In frustrated quantum magnets, long-range magnetic order fails to develop despite a large exchange coupling between the spins. In contrast to the magnons in conventional magnets, their spin excitations are poorly understood. Are they itinerant or localized? Here we show that the thermal Hall conductivity $\kappa_{xy}$ provides a powerful probe of spin excitations in the ``quantum spin ice'' pyrochlore Tb$_2$Ti$_2$O$_7$. The thermal Hall response is large even though the material is transparent. The Hall response arises from spin excitations with specific characteristics that distinguish them from magnons. At low temperature ($T<$ 1 K), the thermal conductivity imitates that of a dirty metal. Using the Hall angle, we construct a phase diagram showing how the excitations are suppressed by a magnetic field.
\end{abstract}
 
\maketitle      
Spin waves or magnons, the elementary excitations in a magnet, can transport heat when a thermal gradient $-\nabla T$ is applied. Because magnons are charge neutral, their thermal Hall conductivity $\kappa_{xy}$ is expected to be zero, i.e. the current is symmetric with respect to sign-reversal of a magnetic field $\bf H$. However, recent findings based on the Berry curvature have overturned this semiclassical result. Following a prediction by Katsura, Nagaosa and Lee (KNL)~\cite{Katsura}, Onose \etal~\cite{Onose} recently observed a weak $\kappa_{xy}$ signal in the ferromagnetic insulator Lu$_2$V$_2$O$_7$. Subsequently, it was pointed out~\cite{Murakami} that $\kappa_{xy}$ should include a contribution from the magnetization current~\cite{Luttinger,Obraztsov,Streda}.

In frustrated magnets, long-range magnetic order fails to develop even at millikelvin temperatures $T$ because of strong quantum fluctuations. The low-lying excitations are currently of great interest~\cite{Molavian,Balents,Gingras}. The pyrochlore Tb$_2$Ti$_2$O$_7$ is an insulator in which Tb and Ti define interpenetrating networks of tetrahedra~\cite{GardnerRMP}. Each Tb$^{3+}$ ion has a large local moment (9.4 Bohr magnetons $\mu_B$), but the lowest lying level is a crystal-field induced spin-$\frac12$ doublet. The spin-spin interaction is antiferromagnetic Ising-type with easy axis along the local (111) axis. Susceptibility experiments report a Curie-Weiss temperature $\theta$ = -19 K~\cite{Gardner1999,GardnerRMP}. However, long-range order is not detected down to 50 mK~\cite{Gardner1999,Gardner2003}. Neutron scattering observed diffuse scattering in zero $H$ which condensed to a Bragg peak at (002) when $H$= 2 T~\cite{Rule}. Recently, neutron diffraction at 50 mK and $H$ = 0 has detected elastic diffusive peaks at $(\frac12,\frac12,\frac12)$ with short correlation length ($\xi\sim 8 \,\rm{\AA}$)~\cite{Ross13} and ``pinch points''~\cite{Fennell}, suggestive of incipient spin-ice order subject to strong quantum fluctuations. Distinct from classical spin-ice pyrochlores (e.g. Dy$_2$Ti$_2$O$_7$)~\cite{Sondhi}, the ground state of Tb$_2$Ti$_2$O$_7$ and Yb$_2$Ti$_2$O$_7$ -- broadly termed ``quantum spin ice''~\cite{Molavian,Balents,Savary2011} -- is predicted to host the quantum spin liquid (at $T=0$) and the thermal spin-liquid (when entropy dominates at finite $T$), as well as a novel Coulomb ferromagnetic state~\cite{Savary2011,Savary2012,Savary2013}. All 3 states harbor exotic excitations. A recent review of the Quantum Spin Ice and Spin Liquid state is given in Ref.~\cite{Gingras}.

In spite of the extensive experimental literature, nearly nothing is known about the transport properties of the spin excitations. Here we show that $\kappa_{xy}$ provides a powerful way to detect the excitations and determine their properties. The observed Hall response in Tb$_2$Ti$_2$O$_7$ becomes quite large below 15 K. The excitations display distinctive characteristics that distinguish them from magnons. At very low $T$, the spin excitations are readily suppressed by $H$, along with the host state. 

We investigated 2 crystals, cut with the $x$-$y$ plane (largest face) normal to (110) in Sample 1 and normal to (111) in 2. With ${\bf H ||\hat{z}}$ and thermal current density $\mathbf{{J}}^q||{\bf\hat{x}}$, we measure the longitudinal gradient $-\partial_x T$ and the transverse gradient $-\partial_y T$ to get the thermal resistivity tensor (defined by $-\partial_i T = W_{ij}J^q_j$), and the thermal conductivity tensor $\kappa_{ij} = W_{ij}^{-1}$ . Throughout, we plot $\kappa_{ij}$ divided by $T$ to remove the entropy factor. The thermal Hall angle is defined by $\tan\theta_H = \kappa_{xy}/\kappa_{xx} = W_{yx}/W_{xx}$. For measurements of $-\partial_y T$, 3 types of thermometers were employed: ruthenium oxide ($T<$ 15 K), Cernox (10$\le T\le$ 40 K) and chromel-constantan thermocouples ($T\ge$ 25 K). The magnetoresistances of the thermometers were extensively calibrated. At 1 K, we resolve 0.5 mK in the Hall signal for a longitudinal $\delta_x T\sim$ 100 mK. The fragility of the crystals above 5 T and the unusually large magneto-caloric effect in Tb$_2$Ti$_2$O$_7$ below 5 K posed challenges which required specially designed mounts and measurement protocols.

\begin{figure*}[t]
\includegraphics[width=12 cm]{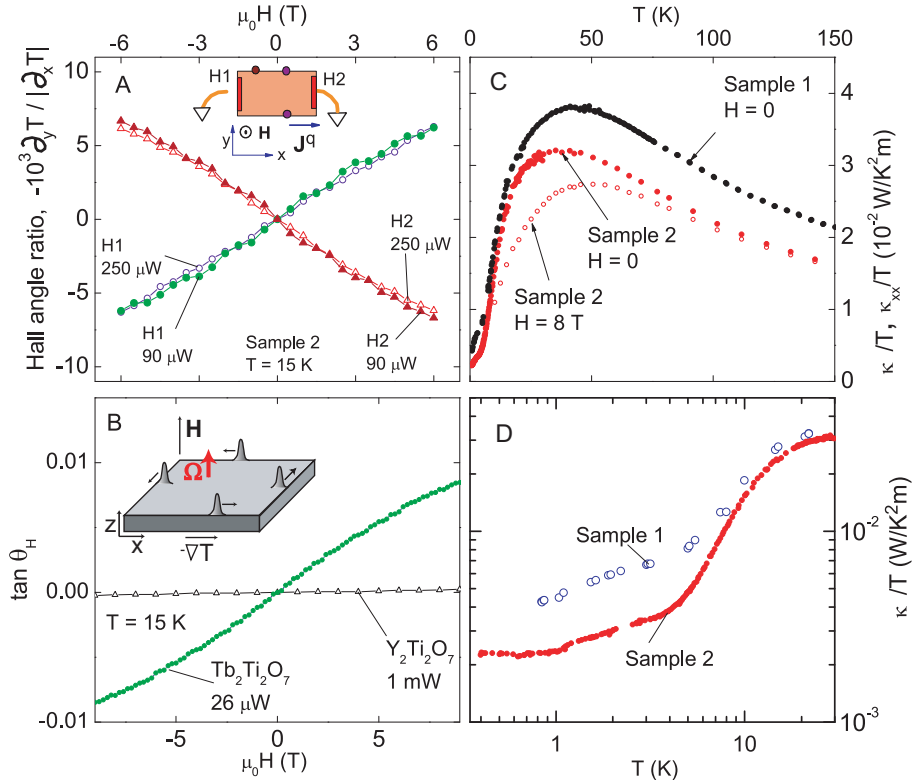}
\caption{\label{figKvsT} 
The thermal Hall effect in the pyrochlore Tb$_2$Ti$_2$O$_7$. Panel (A) shows tests to verify that the large Hall signal is intrinsic. The green solid circles are the Hall angle ratio $-\partial_y T / |\partial_x T|$ measured at 15 K with thermal current density ${\bf J}^q$ flowing to the right (as shown in inset, current applied to heater H1, right edge grounded to bath). When ${\bf J}^q$ points to the left, the Hall angle ratio is inverted in sign (solid triangles) but unchanged in magnitude (to 2 $\%$). When the power is increased 3-fold ($90 \to 270 \,\mu$W), $-\partial_yT / |\partial_x T|$ is nearly unchanged, confirming linear response (open symbols). Panel (B) compares the thermal Hall signal in Sample 1 (solid circles, with applied heater power 26 $\mu$W at 15 K) with the null signal in the non-magnetic analog Y$_2$Ti$_2$O$_7$ with heater power 38$\times$ larger (1 mW at 15 K) (open triangles). The inset shows the wave-packet model proposed by MM~\cite{Murakami} for $\kappa_{xy}$ in a ferromagnetic insulator. Panel (C) plots the $T$ dependence of $\kappa/T$ ($\equiv\kappa_{xx}/T$ at $H=0$) for Samples 1 and 2 (solid symbols). In Sample 2, $\kappa_{xx}/T$ at $H$ = 8 T is also plotted as open circles. The low-$T$ behavior of $\kappa/T$ is shown in Panel (D). In Sample 2, $\kappa/T$ becomes $T$ independent below 1 K, similar to that of a dirty metal.
}
\end{figure*}

Seeing a large thermal Hall signal in an orange transparent crystal is strongly counter-intuitive. Hence we have performed several tests to verify that it is intrinsic. The Hall signal (always hole-like) should be independently antisymmetric in $\bf H$ and in $\mathbf{{J}}^q$. To reverse $\mathbf{{J}}^q$, we warmed up the sample and re-configured the Au wires. The results confirm that the Hall-angle ratio $-\partial_y T/|\partial_x T|$ at 15 K is indeed antisymmetric in both $\bf H$ and $\mathbf{{J}}^q$ (Fig. \ref{figKvsT}A). The ratio is also independent of applied heater power (linear response). In addition, we performed thermal Hall measurements on the non-magnetic analog Y$_2$Ti$_2$O$_7$ (Fig. \ref{figKvsT}B), and verified that it displays a null result (unresolved from our background signal that is $1,700\times$ weaker than in Tb$_2$Ti$_2$O$_7$).

The $T$ dependence of the zero-field thermal conductivity $\kappa$ already reveals an interesting feature at very low $T$. Initially, as we cool below 150 K, $\kappa/T$ attains a broad peak at 40 K and then decreases steeply, reflecting freezing out of the phonons (Fig. \ref{figKvsT}C). Remarkably, below 1 K, $\kappa/T$ settles to a constant value. The constancy in $\kappa/T$, incompatible with phonon or magnon conduction, is reminiscent of a dirty metal. Despite the complete absence of itinerant electrons, the heat current is conveyed by neutral excitations that seem to behave like fermions.

\begin{figure*}[t]
\includegraphics[width=12 cm]{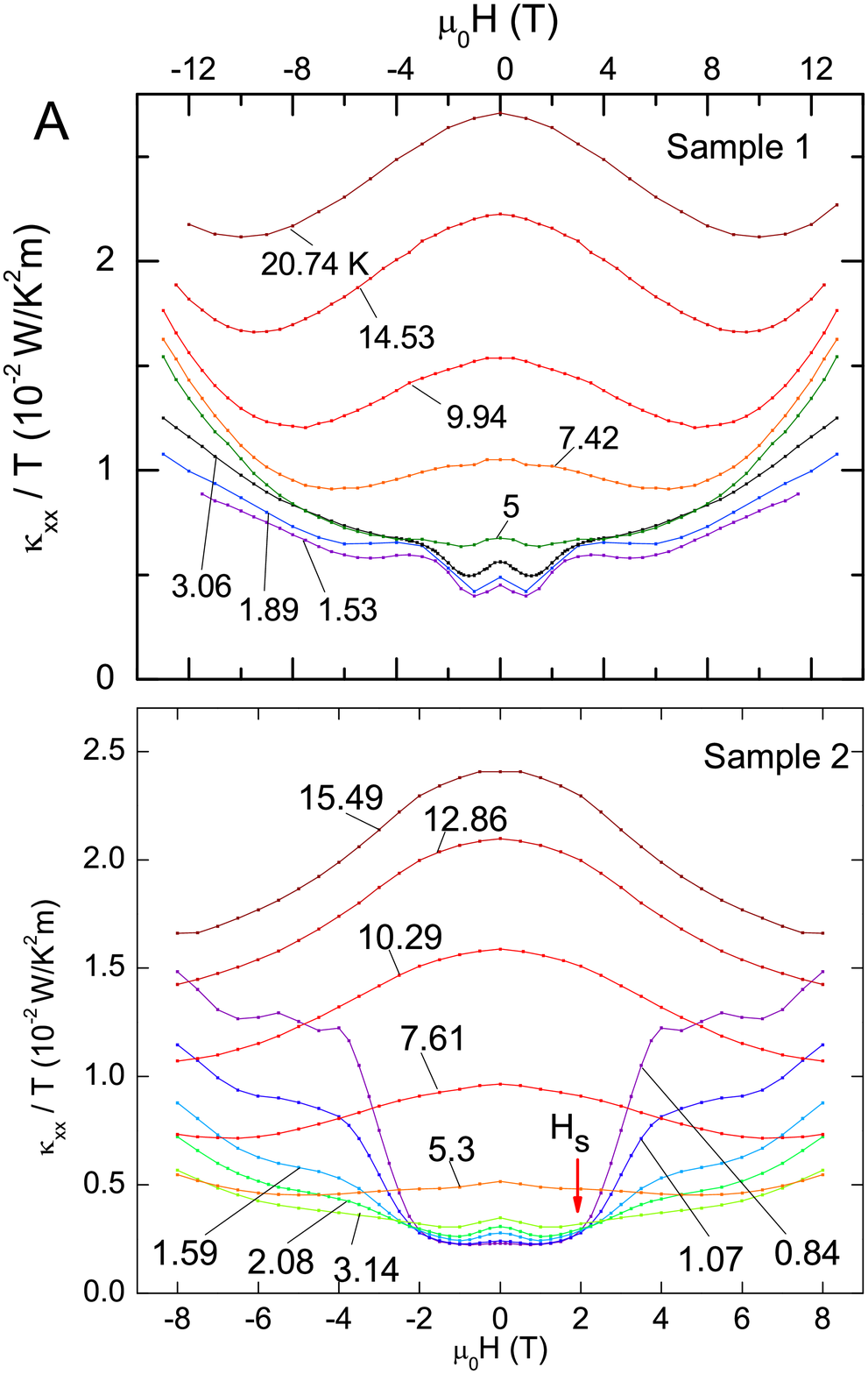}
\caption{\label{figkxx} Curves of $\kappa_{xx}/T$ vs. $H$ in Tb$_2$Ti$_2$O$_7$ in Samples 1 and 2 (Panels A and B, respectively) for $T<$ 21 K. Above 60 K, $\kappa_{xx}/T$ is nearly independent of $H$ (SM). In the interval 5-21 K, the thermal conductivity initially decreases as $H$ increases, but goes through a broad minimum before increasing steeply at larger $H$. Below 5 K, a new feature in Sample 2 becomes apparent in low $H$ (Panel B). At the step-field $H_s\simeq$ 2 T (arrow), $\kappa_{xx}$ undergoes a step increase (by a factor of 4.5 at 0.84 K). 
}
\end{figure*}

In finite $H$, $\kappa_{xx}$ displays the rich pattern of behavior shown in Fig. \ref{figkxx} (see Li \etal~\cite{Sun} for previous measurements of $\kappa_{xx}$ vs. $H$). Above $\sim$80 K, $\kappa_{xx}/T$ is nearly insensitive to $H$, consistent with heat conduction dominated by phonons (Fig. \ref{figkxx}A). Below 80 K, a fairly large $H$ dependence is observed. An 8-T field suppresses $\kappa_{xx}$ by 25-30$\%$ over the broad $T$ interval 7-80 K. In Sample 2, with ${\bf H}||$(111), the field profile is nearly flat for 3$<T<$5 K (Fig. \ref{figkxx}B). Below 3 K, the profile is dominated by a step-increase (by a factor of 4.5 at 0.84 K) that onsets at the step field $H_s$ (arrow). $H_s$ ($\sim$2 T from 0.84 to 2 K) agrees well with the field $H_2$ detected by susceptibility~\cite{Legl}. In neutron scattering~\cite{Rule}, the checkerboard diffuse scattering condenses to a sharp Bragg spot at (002) at 2 T. As seen here, $\kappa_{xx}$ undergoes a 4-fold increase partly from spin waves stiffened by $H$ and from phonon lifetime enhancement [12].

\begin{figure*}[t]
\includegraphics[width=12 cm]{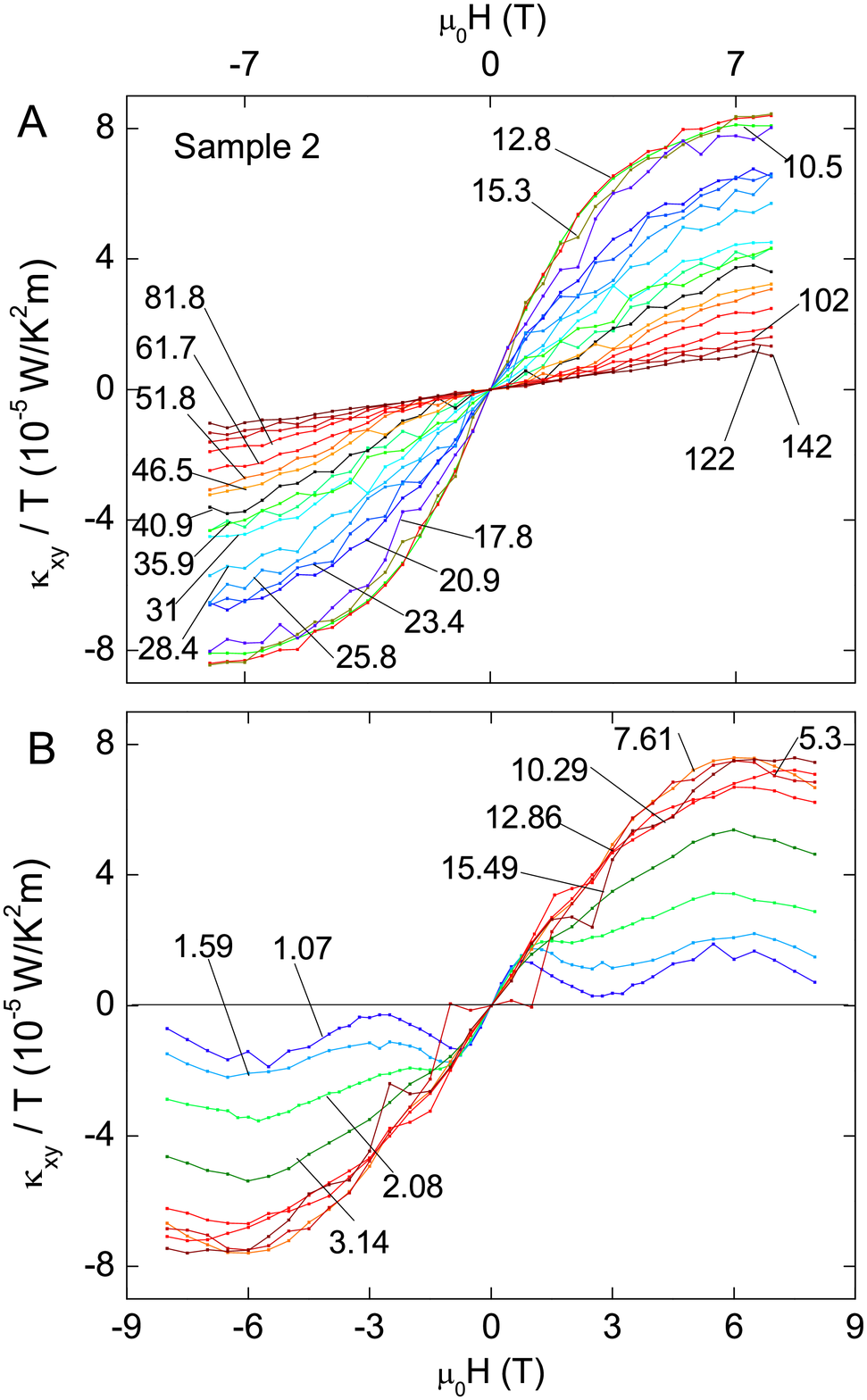}
\caption{\label{figkxy} Curves of the thermal Hall conductivity $\kappa_{xy}/T$ vs. $H$ in Tb$_2$Ti$_2$O$_7$ (Sample 2). From 140 to 50 K, $\kappa_{xy}/T$, is $H$-linear (Panel A). Below 45 K, it develops pronounced curvature at large $H$, reaching its largest value near 12 K. The sign is always ``hole-like''. Panel B shows the curves below 15 K. A prominent feature is that the weak-field slope $[\kappa_{xy}/TB]_0$ is nearly $T$ independent below 15 K. Below 3 K, the field profile changes qualitatively, showing additional features that become prominent as $T\to 0$, namely the sharp peak near 1 T and the broad maximum at 6 T. 
}
\end{figure*}

Figure \ref{figkxy} plots the thermal Hall conductivity (as $\kappa_{xy}/T$) in Sample 2 (the curves for 1 are similar). The Hall signal is positive (``hole-like'') at all $T$. Above 35 K, $\kappa_{xy}/T$ is $H$-linear with a slope that decreases nominally as $1/T$ (Panel A). The thermal Hall response becomes large and generally nonlinear in $H$ below 15 K. Cooling below 3 K reveals several features (Panel B). Close to 1 T, $\kappa_{xy}/T$ displays a shoulder which evolves to a peak feature at 0.84 K. We discuss this important peak in relation to the Hall angle below. Above 1 T, $\kappa_{xy}/T$ goes through a shallow minimum followed by a broad maximum at 6 T. 

A striking feature of the Hall curves in Fig. \ref{figkxy}B is that the initial (weak-field) slope $[\kappa_{xy}/TB]_0$ is nominally $T$-independent below 10 K (where $[\cdots]_0 \equiv \lim_{B\to 0}\cdots$, with $B= \mu_0H$). As $T$ is lowered from 140 K, the rapid increase of $[\kappa_{xy}/TB]_0$ saturates abruptly below 15 K to a constant, within our uncertainties. The thermal Hall effect becomes large below this crossover temperature. Significantly, 15 K is close to the Curie-Weiss scale (-19 K), as well as the excitation gap $\Delta$~\cite{Gingras2000}. 

In contrast, the Hall angle in weak $H$ does not saturate. Inspection of the curves of $\tan\theta_H$ confirms that the initial slope $[\tan\theta_H/B]_0$ continues to increase as $T$ falls from 15 $\to$0.84 K, a behavior also reminiscent of metals, where $[\tan\theta_H/B]_0\sim\tau_{tr}$ (the transport lifetime). As mentioned, $\tan\theta_H$ displays a prominent peak at the ``peak field'' $H_p(T)$ (see arrow). For $H>H_p$, the Hall response is strongly suppressed (see especially the curves at 1 K and 0.84 K). We note that $H_p$ nearly coincides with the low field-scale $H_1$ (dashed curve) recently reported by Legl \etal~\cite{Legl}.

Further information on the crossover field comes from the Hall angle slope $\tan\theta_H/B$. It is useful to regard $\tan\theta_H/B$ as a ``susceptibility'' that measures the strength of the off-diagonal (Hall) response. At zero $H$, $\tan\theta_H/B$ rises monotonically as $T\to$ 0.84 K. At finite $H$, however, each curve displays a distinct maximum which defines the field scale $H'_p(T)$. Plotted in Panel B, the curves of $H'_p$ (open circles) and $H_p$ (solid) demarcate the region in which the Hall susceptibility $\tan\theta_H/B$ is large (shown shaded). The ``phase diagram'' highlights a crucial feature of the Hall excitations and its host quantum state. The Hall susceptibility $\tan\theta_H/B$ peaks at $H_p$, and falls steadily reaching nearly zero at the step field $H_s$. To us, this suggests that the host state, subject to strong quantum fluctuations, is readily suppressed by a large $H$. At low $T$, it is confined to the narrow wedge defined by the field $H_p$ (and $H_p'$). We discuss next why $\kappa_{xy}$ cannot be from phonons or magnons.

A weak phonon Hall effect was observed~\cite{Strom} in the garnet Tb$_3$Ga$_5$O$_{12}$. We already mentioned that the constancy of $\kappa/T$ below 1 K in our samples (Fig. \ref{figKvsT}C) is incompatible with phonon conduction. Despite the small $\kappa$, $\tan\theta_H$ (at 1 T) is 90$\times$ larger than in the garnet. More importantly, the curve of $\tan\theta_H$ vs. $H$ provides a sharp test. In the phonon scenario, skew scattering from local moments in the disordered state yields a weak Hall signal at low $H$. As $H_s$ increases, increased alignment of the moments should lead to an increasing $\tan\theta_H$, especially above $H_s$ where the magnetization is $\sim 3\mu_B$ per Tb ion~\cite{Legl}. Instead, the opposite is observed; $\tan\theta_H$ rapidly falls to nearly zero as the host quantum state is suppressed above $H_p$. This strong contradiction persuades us that $\kappa_{xy}$ here is not from phonons.

The true ground state of Tb$_2$Ti$_2$O$_7$ is still elusive~\cite{Gingras}. The incipient two-in, two-out correlation measured at 70 mK~\cite{Ross13} has a very short $\xi$ ($\sim 8 \,\rm\AA$). At our $T$ ($>$0.84 K), $\xi$ should be even shorter, so the spin excitations responsible for $\kappa_{xy}$ cannot be conventional magnons. Indeed, the constancy of $[\kappa_{xy}/T]_0$ below 15 K in Fig. \ref{figkxy}C violates the power-law $T^a$ dependence (with $a>2$) predicted by the magnon model~\cite{Murakami}.

Instead, our results point to \emph{neutral} excitations subject to a novel kind of Lorentz force ${\bf F}_{\rm L} =e_s \bf v\times B$, with $e_s$ an effective charge and $\bf v$ the drift velocity driven by $-\nabla T$ (Fig. \ref{figKvsT}A). As noted by KNL~\cite{Katsura}, if the spin excitations are fermionic, the Wiedemann-Franz law requires $\kappa/T$ to be $T$-independent at low $T$. Hence the constancy of $\kappa/T$ seems consistent with neutral fermionic excitations below 1 K where mean-free-path is no longer $T$ dependent. From the sign of $\kappa_{xy}$, $e_s$ is positive. A crude estimate gives $e_s$ = 70-540$\times$ the elemental charge $e$.


\vspace{1cm}\noindent
$^*$Corresponding author. E-mail: npo@princeton.edu

\vspace{1cm}\noindent
{\bf Acknowledgements}\\
We acknowledge discussions with P. A. Lee, S. Murakami and N. Nagaosa. The research is supported by the Army Research Office (ARO W911NF-11-1-0379, ARO W911NF-12-1-0461) and the US National Science Foundation (grant DMR 0819860). The growth of the crystal and characterization were performed by J.W.K. and R.J.C. with the support of the DOE, Division of Basic Energy Sciences, grant DE-FG-02-08ER46544.

\newpage

\newpage



\begin{thebibliography}{99}

\bibitem{Katsura} H. Katsura, N. Nagaosa and P. A. Lee,
``Theory of the Thermal Hall Effect in Quantum Magnets,"
Phys Rev Lett {\bf 104}, 066403 (2010).

\bibitem{Onose} Y. Onose, T. Ideue, H. Katsura, Y. Shiomi, N. Nagaosa and Y. Tokura,
``Observation of the Magnon Hall Effect,'' 
Science {\bf 329}, 297-299 (2010).

\bibitem{Murakami} R. Matsumoto and S. Murakami,
``Theoretical Prediction of a Rotating Magnon Wave Packet in Ferromagnets,'' 
Phys Rev Lett {\bf 106}, 197202 (2011).


\bibitem{Luttinger} J. M. Luttinger,
``Theory of Thermal Transport Coefficients,''
Phys. Rev. {\bf 135}, A1505 (1964).

\bibitem{Obraztsov} Y. N. Obraztsov,
``Thermomagnetic Phenomena in Metals and Semiconductors in Quantizing (Strong) Magnetic Fields,'' 
Sov Phys-Sol State {\bf 6}, 331-336 (1964).

\bibitem{Streda} H. Oji and P. Streda,
``Theory of Electronic Thermal Transport - Magnetoquantum Corrections to the Thermal Transport-Coefficients,''
Phys Rev B {\bf 31}, 7291-7295 (1985).

\bibitem{Molavian} Hamid R. Molavian, Michel J. P. Gingras and Benjamin Canals,
``Dynamically Induced Frustration as a Route to a Quantum Spin Ice State in Tb$_2$Ti$_2$O$_7$
via Virtual Crystal Field Excitations and Quantum Many-Body Effects,''
Phys. Rev. Lett. {\bf 98}, 157204 (2007). DOI: 10.1103/PhysRevLett.98.157204

\bibitem{Balents} L. Balents,
``Spin liquids in frustrated magnets,'' 
Nature {\bf 464}, 199-208 (2010).


\bibitem{Gingras} M. J. P. Gingras and P. A. McClarty,
``Quantum spin ice: a search for gapless quantum spin liquids in pyrochlore magnets,''
Rep. Prog. Phys. {\bf 77} 056501 (2014), doi:10.1088/0034-4885/77/5/056501


\bibitem{GardnerRMP} J. S. Gardner, M. J. P. Gingras and J. E. Greedan,
``Magnetic pyrochlore oxides,''
Rev Mod Phys {\bf 82}, 53-107 (2010).

\bibitem{Gardner1999} J. S. Gardner, S. R. Dunsiger, B. D. Gaulin, M. J. P. Gingras, J. E. Greedan, R. F. Kiefl, M. D. Lumsden, W. A. MacFarlane, N. P. Raju, J. E. Sonier, I. Swainson and Z. Tun,
``Cooperative paramagnetism in the geometrically frustrated pyrochlore antiferromagnet Tb$_2$Ti$_2$O$_7$,'' 
Phys Rev Lett {\bf 82}, 1012-1015 (1999).


\bibitem{Gardner2003} J. S. Gardner, A. Keren, G. Ehlers, C. Stock, E. Segal, J. M. Roper, B. Fak, M. B. Stone, P. R. Hammar, D. H. Reich and B. D. Gaulin,
``Dynamic frustrated magnetism in Tb$_2$Ti$_2$O$_7$ at 50 mK,'' 
Phys Rev B {\bf 68}, 180401 (2003).

\bibitem{Rule} K. C. Rule, J. P. C. Ruff, B. D. Gaulin, S. R. Dunsiger, J. S. Gardner, J. P. Clancy, M. J. Lewis, H. A. Dabkowska, I. Mirebeau, P. Manuel, Y. Qiu and J. R. D. Copley,
``Field-induced order and spin waves in the pyrochlore antiferromagnet Tb$_2$Ti$_2$O$_7$,'' 
Phys Rev Lett {\bf 96}, 177201 (2006).




\bibitem{Ross13} K. Fritsch, K. A. Ross, Y. Qiu, J. R. D. Copley, T. Guidi, R. I. Bewley, H. A. Dabkowska and B. D. Gaulin,
``Antiferromagnetic spin ice correlations at (1/2, 1/2, 1/2) in the ground state of the pyrochlore magnet Tb$_2$Ti$_2$O$_7$,''
Phys. Rev. B {\bf 87}, 094410 (2013).


\bibitem{Fennell} T. Fennell, M. Kenzelmann, B. Roessli, M. K. Haas, and R. J. Cava,
``Power-law spin correlations in the pyrochlore antiferromagnet Tb$_2$Ti$_2$O$_7$,''
Phys. Rev. Lett. {\bf 109}, 017201 (2012).


\bibitem{Sondhi} C. Castelnovo, R. Moessner and S. L. Sondhi, 
``Spin ice, fractionalization, and topological order,''
Annual Rev. Cond. Matter Phys. {\bf 3}, 35-55 (2012).


\bibitem{Savary2011} K. A. Ross, L. Savary, B. D. Gaulin and L. Balents,
``Quantum Excitations in Quantum Spin Ice,''
Phys Rev X {\bf 1}, 021002 (2011).

\bibitem{Savary2012} Lucile Savary and Leon Balents, 
``Coulombic Quantum Liquids in Spin-$\frac12$ Pyrochlores,''
Phys. Rev. Lett. {\bf 108}, 037202 (2012).

\bibitem{Savary2013} Lucile Savary and Leon Balents, 
``Spin liquid regimes at nonzero temperature in quantum spin ice,''
Phys. Rev. B {\bf 87}, 205130 (2013).


\bibitem{Sun} Q. J. Li, Z. Y. Zhao, C. Fan, F. B. Zhang, H. D. Zhou, X. Zhao and X. F. Sun,
``Phonon-glass-like behavior of magnetic origin in single-crystal Tb$_2$Ti$_2$O$_7$,'' 
Phys Rev B {\bf 87}, 214408 (2013).


\bibitem{Legl} S. Legl, C. Krey, S. R. Dunsiger, H. A. Dabkowska, J. A. Rodriguez, G. M. Luke and C. Pfleiderer,
``Vibrating-Coil Magnetometry of the Spin Liquid Properties of Tb$_2$Ti$_2$O$_7$,'' 
Phys Rev Lett {\bf 109}, 047201 (2012).




\bibitem{Gingras2000} M. J. P. Gingras, B. C. den Hertog, M. Faucher, J. S. Gardner, S. R. Dunsiger, L. J. Chang, B. D. Gaulin, N. P. Raju and J. E. Greedan,
``Thermodynamic and single-ion properties of Tb3+ within the collective paramagnetic-spin liquid state of the frustrated pyrochlore antiferromagnet Tb$_2$Ti$_2$O$_7$,''
Phys Rev B {\bf 62}, 6496-6511 (2000).



\bibitem{Strom} C. Strohm, G. L. J. A. Rikken, and P. Wyder,
``Phenomenological Evidence for the Phonon Hall Effect,''
Phys. Rev. Lett. {\bf  95}, 155901 (2005).

\end{thebibliography}
\end{document}